# Outcome orientation —

# a misconception of probability that harms medical research and practice

Parris T. Humphrey[a], Joanna Masel[a,*]

[a]Department of Ecology & Evolutionary Biology

University of Arizona

1041 E Lowell St.

Tucson, AZ. 85721, USA

masel@email.arizona.edu, pth@email.arizona.edu

*Corresponding author Ph. 520 626 9888

**Keywords:** randomized controlled trial; cognitive bias; medical practice; personalized medicine; N of 1; probability

## Abstract

Human understanding of randomness and variation is shaped by a number of cognitive biases. Here we relate a lesser-known cognitive bias, the "outcome orientation", to medical questions and describe the harm that the outcome orientation can do to medical research and practice. People who reason according to an outcome orientation interpret probability as a subjective degree of belief that is constrained to consider events one at a time, in a way that is incompatible with Bayesian reasoning. Instead of accepting that uncertainty is inevitable, and generalizing from the frequency of similar events, the outcome orientation prefers one-off causal narratives. In medicine, the outcome orientation therefore erodes support for randomized controlled trials in favor of reductionist approaches. The rhetoric of personalized medicine resonates with, and can promote, the outcome orientation, by emphasizing how the measurable attributes of individual patients, rather than chance or unknowable factors, causally produce each particular patient's outcome.

*"We are far too willing to reject the belief that much of what we see in life is random."*

–D. Kahneman (2011)

Uncertainty is an everyday experience in medical research and practice, but theory and methods for reasoning clearly about uncertainty were developed only recently (Doll 1998, Hacking 1984, Salsburg 2001). Confirmation bias, selective memory, and many misleading heuristics are known enemies of the insightful clinician, researcher, or citizen (Kahneman 2011); but other snares worth exposing may lurk in how we reason about uncertainty in our everyday lives. Here we draw attention to a cognitive bias described by Konold as the "outcome orientation" (Konold 1989) — little known or possibly unknown to those outside the field of probability pedagogy — and point out how this form of reasoning creates hazards for medical research and practice.

Imagine a weather forecast of 70% rain. Most people round this up to "it will rain". If it doesn't, you curse the forecaster for making a mistake (if you have cancelled your picnic). Now imagine 10 days, all of which had a 70% rain forecast. The most likely outcome, given an accurate forecaster, is that it will rain on 7 out of the 10 days. But nearly half the students beginning our University of Arizona Evidence-Based Medicine class believed that if a weather forecaster were good at his job, it should rain on all 10 days (Masel et al. 2015), in line with previous studies of statistics classes (Konold 1995). People with an outcome orientation do not interpret probability statements in terms of the expected frequency of an outcome when a task is repeated many times. Instead, their "goal in dealing with uncertainty is to predict the outcome of a single next trial" (Konold 1989). Asked to predict the next 6 rolls of a dice with 5 black sides and one white, someone thinking in terms of an outcome orientation fails to consider the 6 rolls as a group, and so believes that the most likely outcome is 6 rolls of black, rather than 5 black rolls and one white.

An outcome orientation also changes the way that people value evidence. Konold gave his students an irregularly shaped bone and asked them to estimate which of six sides would most likely land upright if the bone were rolled like a dice (Konold 1989). Students with an outcome orientation were happy to call their estimate as right or wrong based on a single roll and were remarkably uninterested in gathering data on multiple rolls, or being given data on the outcomes of 1000 previous bone rolls. Instead, they preferred to be given information on the surface area of each side, or a technical drawing showing the center of gravity, despite the self-evident impracticality of converting this information into reliable predictions. Despite the fact that it is straightforward to convert data on previous rolls into a prediction, by simply choosing the most common outcome, some felt that previous rolls, unlike physical properties, were not "real evidence".

In a clinical setting, the treatment of one patient is like the roll of one dice. Randomized controlled trials (RCTs) summarize the results of many patients/dice rolls, comparing the outcomes between groups of patients who are assigned to different treatments. While each patient is of course unique, both the known and the unknown variables underlying that uniqueness are equally represented in the different arms of the RCT, to within a degree of precision effectively summarized by statistical analysis. Because the randomization of interventions controls for both known and unknown confounding factors, RCTs provide uniquely reliable information about which treatments are effective for which diseases, and are responsible for dramatic improvements in medicine (Burch 2009, Doll 1998). Unfortunately, the statistical nature of RCT findings—based on aggregates and not individuals—is persistently unpopular among patients and doctors (Burch 2009). Backlash against RCTs puts continued medical progress at risk.

The outcome orientation helps explain the nature of psychological resistance to RCT evidence, information that may help in defending the role of RCTs. Patients abhor uncertainty about their own outcome, and the outcome orientation, unlike a full understanding of RCTs, does not demand that they accept such uncertainty as inevitable. What is more, patients aren't interested in whether a particular treatment is the most successful for the average person with their condition, the information that an RCT provides. They want to know whether the treatment will work for them. Medical decisions are necessarily made with a focus on "the outcome of a single next trial", namely that of the current patient, rather than a focus on a group of patients. This is true both for the patient ("will this work?") and for the medical provider ("what treatment should I choose this time?"). This framing of the question in terms of a single trial, rather than a group of trials, predisposes patients and providers to reason according to an outcome orientation. Each patient has only one life; the patient is Konold's bone, and the doctor metaphorically "rolls" the patient once or several times only.

Patient variability is undeniable. But an outcome orientation changes the lessons that patients and doctors draw from this fact. The message from the science of probability is that RCTs are the best way, in an uncertain world, to determine the optimal treatment for a patient given a particular diagnosis, by generalizing from others with the same diagnosis who were randomized to treatments in order to balance out the effects of confounding factors. Treatments are evaluated based on statistical trends in the face of variation. When a doctor does *not* suffer from the outcome orientation, they can use RCT data to reason about the optimal prescribing policy in order to maximize the total wellbeing of all patients in their practice, and then follow this policy when faced with individual patients. In contrast, use of the outcome orientation leads a doctor to focus on hard to interpret individual patient idiosyncracies (the technical drawing of the bone), to the point of substantial deviation from the RCT evidence (the history of previous rolls of similar bones).

For example, following a recent RCT indicating that steroid injections are ineffective for spinal stenosis (Friedly et al. 2014), The New York Times reported Dr. Thiru Annaswamy of the Dallas Veterans Medical Center as denying this evidence by saying that "there are patients who clearly respond to steroid injection. However, it is unclear why some do, and others don't" and that a particular patient "may have responded to the lidocaine-only [control] injection" (Belluck 2014). This indicates a deep-seated belief, shared by doctors and the media as well as by patients, that all outcomes have causes. Rather than accept that some outcomes are intrinsically random, all patient improvements and deteriorations are seen as "responses", whether the response is to the treatment or to the active placebo.

Dr. Annaswamy may be correct that if the experiment could be repeated, certain patients would consistently have better outcomes with the treatment and others with the active placebo. But even if he were correct (an assertion for which we have as yet insufficient evidence), one measurement per patient would not be enough to know which patient is which. After all, the unique makeup of individual patients is responsible for only some of the variation around those trends. Random fluctuations in disease severity *in the same patient* over time can also make it look like some but not all patients "respond" to a treatment. We currently do not understand why each patient outcome arises the way it does. In the absence of such understanding, variation among patient outcomes is best treated as random, via a RCT. An outcome orientation increases the temptation not only to intuit that causes exist, but also to act as though a single measurement per patient were sufficient to infer them.

Unfortunately, just as subjects in Konold's study were uninterested in rolling a bone multiple times in order to predict future rolls, doctors as well as patients with an outcome orientation may downplay the results of RCTs and look instead to more personalized[1] alternatives. Personalized medicine resonates with patients, and the urge to offer treatment recommendations that appear tailored to a specific patient may sway the decisions of many doctors. Individual patients, and their doctors, may see individual experiences as sufficient evidence to justify these personalized treatment choices, even in the face of contradictory RCT data. For example, during the 2005 FDA hearings that helped determine the future of cox-2 inhibitor availability, patients gave moving testimonials about how only rofecoxib (Vioxx®) or celecoxib (Celebrex®) gave them pain relief, and no other drug (Center for Drug Evaluation and Research 2005). An outcome orientation leads doctors and patients to overweight the claims of unsystematic and un-statistical "works for me" forms of anecdotal evidence. Even when doctors know better, a doctor's rejection of outcome orientation reasoning can be difficult to adhere to when patients state their claims with conviction.

---

[1] Here we define "personalized medicine" as "treatment decisions intended to vary among patients with the same diagnosis".

The hypothesis that the regular clinical experience of a single patient contains sufficient information to guide future treatment can be rigorously tested using a formal "N of 1" study design. Patients who believe that they respond uniquely to cox-2 inhibitors would enroll in a formal trial to receive a randomized double-blind sequence of cox-2 inhibitors interspersed with an alternative medication, and report their pain. This study design would provide information regarding both the strength of such personalized claims as a class, and their likely validity for particular patients. It is telling that this formal N of 1 approach is rarely pursued, despite its suitability for testing common claims of the "it works for me" type. This may be explained by the fact that the outcome orientation undermines not just RCTs that average across patients, but also N of 1 RCTs that average across time points (i.e. dice rolls) for the *same* patient. N of 1 trials can effectively quantify the usefulness of using small amounts of individual-based data to personalize the treatment of certain chronic conditions (Guyatt et al. 1986); if successful, they can provide medical decisions that are simultaneously personalized, and also evidence-based (Lillie et al. 2011).

Despite these benefits, formal and rigorously conducted N of 1 studies remain rare (Lillie et al. 2011). We argue that this is in part because they are subject to the same rejection of modern notions of uncertainty, probability, and frequency as other RCTs. After all, their personalized nature should remove some of the objections to standard RCTs, and so create an opportunity for patients and their doctors to grasp the role of random fluctuations in their personal "outcomes". The lack of interest in well-designed N of 1 studies, despite their feasibility for many conditions even on a small scale, suggests that rejection of intrinsic randomness (associated with an outcome orientation) may be a bigger obstacle to RCTs than lack of personalization. Identifying the psychological or cognitive factors that undermine the perceived value of population-level or individual RCTs is crucial in order to ensure that medical approaches, whether personalized or not, proceed with a sufficient evidence base.

Just as Konold's students thought they could learn more from the physical characteristics of the bone than they could learn from past rolls, many patients seek to open the black box of their diagnosis and learn how their unique physical characteristics shape their individual risks. Personalized and patient-centered medicine calls for renewed appreciation of individual variation, and modern biomedical research has been quick to answer this call by seeking the mechanistic and statistical causes of variation in disease outcomes (Hood et al. 2004). Although some patients do have very different baseline risks or respond differently to different treatments for biologically causal reasons (e.g. HER2 alleles for breast cancer, CCR5-$\Delta$32 alleles for HIV susceptibility), results from RCTs on diagnostics show that having more data only rarely helps patients (Siontis et al. 2014). Furthermore, expecting causal explanations from precise measurements fuels the dangerous argument that –omics derived biomarkers are justified by such a compelling causal theory that they do not require RCT-level validation prior to implementation (Frueh 2009). This position requires continued pushback, given our lamentable history of vainly predicting successful results from phase III trials (Ioannidis and Khoury 2013, Ioannidis and Khoury 2011). Unlike the sale of new drugs, the implementation of decision-making based on available biomarkers does not require regulatory approval, and so the eventual availability of RCT data cannot be taken for granted.

Tellingly, Darwin wrote that he had "*hitherto sometimes spoken as if the variations... had been due to chance. This, of course, is a wholly incorrect expression, but it serves to acknowledge plainly our ignorance of the cause of each particular variation*" (Darwin 1859). Even in a deterministic universe, the causes of particular variations may be not merely unknown, but fundamentally unknowable. The promise of personalized medicine to uncover the cause of each particular variation may therefore be treacherous; accepting instead the inevitability of randomness and uncertainty leads, ironically, to greater knowledge via the acceptance of the primacy of RCTs.

RCTs, through their use of a rigorous control group, correct not only for variation across patients and variation in the same patient across time, but also for placebo effects, including regression to the mean and reporting biases. These attributes of rigorous RCT-derived evidence help buffer doctors from being overly influenced by individual patient outcomes.

Even when the gold-standard internal validity of RCTs is acknowledged on these grounds, RCTs can be criticized on the basis of their external validity, i.e. that the patients enrolled in the RCT may not be representative of the patients who will eventually be prescribed the treatment. Clay (2010) offers an example of such a legitimate critique in the American Psychological Association's magazine. However, this author simultaneously objects that RCTs "don't tell you the critical information you need, which is which patients are going to benefit from the treatment", as though this were a reason to use other forms of evidence instead of RCTs. This indeed may be the thing that clinicians and their patients would benefit most from knowing. But the belief that this can be known, with or without evidence from RCTs, reflects a rejection of the inevitability of uncertainty, phrased in a way that is typical for an outcome orientation. Instead of proposing that we solve the external validity problem of some RCTs—by enrolling a more representative set of patients—the author instead suggests that the problem would be solved by using non-RCT forms of evidence. For example, a phase II non-randomized trial was cited as an example of alternative study design, despite the obvious fact that its external validity is no better than a RCT. Similarly, multiple observations of the same subject was presented as an alternative to RCTs, as though it were incompatible with randomization, whereas in fact it is central to the randomized N of 1 study design. Inconsistent critiques such as that of Clay (2010) are best explained as post-hoc rationalizations of an aversion to RCTs driven by the outcome orientation.

The harms of the outcome orientation are many. An outcome orientation can convince patients that, because they truly are "not a statistic", evidence from RCTs does not apply to them. It can convince doctors that they have failed a patient who does badly despite receiving the statistically supported treatment, or that the treatment itself is no good. It can persuade us all that variation in the way doctors treat disease, in the absence of evidence and outside of a RCT, is not merely normal and defensible, but even admirable and a sign of wisdom and special expertise. And it can divert research resources away from RCTs (including rigorous N of 1 study designs) and towards molecular reductionist approaches that do less to improve patient outcomes (Horrobin 2003). Konold's subjects were reluctant to roll their bones to collect and use statistical data; they turned their backs on good evidence in favor of a quest for mechanistic understanding derived from precise measurements.

Given the prevalence of the outcome orientation among Konold's students, we expect the outcome orientation to be common among the general patient population. Given the way in which the framing of medical decisions primes the outcome orientation, we also expect it to be at play among pre-medical and medical students. We can certainly see evidence of the outcome orientation among medical journalists and the doctors they write about (Belluck 2014). Combatting this form of reasoning is crucial for fostering excellence in medical practice, as well as helping patients understand their own health and treatment options. Rather than relying on unreliable intuitions such as the outcome orientation, doctors and patients alike require interventions to develop their intuitions about randomness.

Research is desperately needed on such interventions to combat the outcome orientation at every level of education, from K–12 to continuing medical education. Testing the mostly pre-health undergraduates both before and after taking our own Evidence-Based Medicine class (Masel et al. 2015), we saw an improvement from 3.20 to 2.35 ($p < 0.05$, paired t-test), which corresponds to 17 out of our 40 students giving one fewer outcome oriented response by the end of the course, on the mean "outcome orientation misconception" sub-score of the Quantitative Reasoning Quotient (Sundre 2003) tool, comparable to the 6 percentage point improvement on the 70% rain question that was seen in a previous study (Konold 1995). This is encouraging because it suggests that change may be possible. But in order to make a difference, short-term measures of improved reasoning about uncertainty must carry through over the long-term, into real world scenarios that occur every day in medical settings. Developing reliable metrics of reasoning that predict downstream behavior remains a worthy challenge for evaluators, helping instructors set goals for their educational interventions. However, in order for the seriousness of this call to be heard, the harms of the outcome orientation must first be more widely acknowledged.

In the context of medicine, the "outcome approach" to probability implicitly rejects the entire rationale for randomized controlled trials by focusing only on predicting individual outcomes and failing to consider frequencies as a property of a group. The outcome orientation undermines support for rigorous statistical forms of evidence in favor of anecdotal individual outcomes and experiences and a potentially fruitless quest for personalized causes. The goals of personalized and genomic medicine resonate with, and can even promote, an outcome approach to probability if motivated by implicit beliefs that each variant in outcome has an explainable cause. Articulations of the outcome orientation in the media underscore the prevalence of the problem amongst the otherwise well educated.

It may be easy to shrug off the outcome orientation in the context of rolling a six-sided bone or dice. Perhaps it feels more tempting with a 70% rain forecast than a six-sided bone. In medicine, the stark contrast between the individuality of a patient and the nature of statistical evidence makes an outcome orientation feel natural, but the outcome orientation is just as fallacious for patients as it is for the rolls of a dice. Unless we learn to recognize and counter this fallacious mode of reasoning, it could undermine the power and potential of modern medicine.

**Acknowledgements:** This work was funded in part by a grant to the University of Arizona from the HHMI (52006942). The opinions expressed herein do not necessarily represent those of our funders, who played no role in the preparation of this article.

## References:

Belluck, Pam. 2014. "Common Back and Leg Pain Treatment May Not Help Much, Study Says." *The New York Times*, July 3.

Burch, Druin. 2009. *Taking the Medicine: A Short History of Medicine's Beautiful Idea, and Our Difficulty Swallowing It*. London: Chatto & Windus.

Center for Drug Evaluation and Research. 2005. Joint meeting of the arthritis advisory committee and the drug safety and risk management committee. http://www.fda.gov/ohrms/dockets/ac/05/transcripts/2005-4090T2.htm.

Clay, Rebecca A. 2010. "More than one way to measure." *Monitor on Psychology* 41 (8): 52-55.

Darwin, Charles Robert. 1859. "Origin of Species." p. 131.

Doll, Richard. 1998. "Controlled trials: the 1948 watershed." *Br Med J* 317 (7167): 1217-1220.

Friedly, Janna L., Bryan A. Comstock, Judith A. Turner, Patrick J. Heagerty, Richard A. Deyo, Sean D. Sullivan, Zoya Bauer, Brian W. Bresnahan, Andrew L. Avins, Srdjan S. Nedeljkovic, David R. Nerenz, Christopher Standaert, Larry Kessler, Venu Akuthota, Thiru Annaswamy, Allen Chen, Felix Diehn, William Firtch, Frederic J. Gerges, Christopher Gilligan, Harley Goldberg, David J. Kennedy, Shlomo Mandel, Mark Tyburski, William Sanders, David Sibell, Matthew Smuck, Ajay Wasan, Lawrence Won, and Jeffrey G. Jarvik. 2014. "A Randomized Trial of Epidural Glucocorticoid Injections for Spinal Stenosis." *N Engl J Med* 371 (1): 11-21.

Frueh, Felix W. 2009. "Back to the future: why randomized controlled trials cannot be the answer to pharmacogenomics and personalized medicine." *Pharmacogenomics* 10 (7): 1077-1081.

Guyatt, Gordon, David Sackett, D. Wayne Taylor, John Ghong, Robin Roberts, and Stewart Pugsley. 1986. "Determining Optimal Therapy — Randomized Trials in Individual Patients." *N Engl J Med* 314 (14): 889-892.

Hacking, Ian. 1984. *The Emergence of Probability: A Philosophical Study of Early Ideas about Probability, Induction and Statistical Inferenc*, *Cambridge Series on Statistical & Probabilistic Mathematics*: Cambridge University Press.

Hood, Leroy, James R Heath, Michael E Phelps, and Biaoyang Lin. 2004. "Systems biology and new technologies enable predictive and preventative medicine." *Science* 306 (5696): 640-643.

Horrobin, David F. 2003. "Modern biomedical research: an internally self-consistent universe with little contact with medical reality?" *Nat Rev Drug Discov* 2 (2): 151-154.

Ioannidis, John, and Muin Khoury. 2013. "Are randomized trials obsolete or more important than ever in the genomic era?" *Genome Med* 5 (4): 32.

Ioannidis, John P. A., and Muin J. Khoury. 2011. "Improving Validation Practices in "Omics" Research." *Science* 334 (6060): 1230-1232.

Kahneman, D. 2011. *Thinking, Fast and Slow*. New York: Farrar, Straus and Giroux.

Konold, Clifford. 1989. "Informal Conceptions of Probability." *Cognition and Instruction* 6 (1): 59-98.

Konold, Clifford. 1995. "Issues in assessing conceptual understanding in probability and statistics." *Journal of Statistics Education* 3 (1): 1-9.

Lillie, Elizabeth O, Bradley Patay, Joel Diamant, Brian Issell, Eric J Topol, and Nicholas J Schork. 2011. "The n-of-1 clinical trial: the ultimate strategy for individualizing medicine?" *Personalized medicine* 8 (2): 161-173.

Masel, Joanna, Parris T. Humphrey, Brenna Blackburn, and Joshua A Levine. 2015. "Evidence-based medicine as a tool for undergraduate probability and statistics education." *CBE - Life Science Education* in press.

Salsburg, David. 2001. *The lady tasting tea: How statistics revolutionized science in the twentieth century*: Macmillan.

Siontis, Konstantinos C., George C. M. Siontis, Despina G. Contopoulos-Ioannidis, and John P. A. Ioannidis. 2014. "Diagnostic tests often fail to lead to changes in patient outcomes." *J Clin Epidemiol* 67 (6): 612-621.

Sundre, Donna L. 2003. Assessment of Quantitative Reasoning to Enhance Educational Quality. In *American Educational Research Association meeting*. Chicago, Illinois www.stat.auckland.ac.nz/~iase/cblumberg/sundreqrqpaper.pdf.